\documentclass[10pt, a4paper, twocolumn, aps, pra, showpacs, longbibliography, nofootinbib,superscriptaddress]{revtex4-2}%
\usepackage{mystyle}
\usepackage{graphicx}
\usepackage{dcolumn}
\usepackage{bm}
\usepackage{upgreek}
\def\rd{r_{{\rm d}}}
\newcommand{\COMMENTED}[1]{}

\definecolor{purple}{rgb}{0.5, 0.0, 0.5}

\definecolor{gold}{rgb}{0.83, 0.69, 0.22}

\usepackage[normalem]{ulem}

\def\eevar{\mathfrak{e}^2}
\def\rsvar{\mathfrak{r}_{\text{s}}}
\def\aB{a_{\text{B}}}
\def\rs{r_s}
\def\rbf{{\bf r}}

\begin{document}

\preprint{APS/123-QED}

\title{Critical gate distance for Wigner crystallization in the two-dimensional electron gas}

\author{Agnes~Valenti}
\thanks{A.V. and V.C. contributed equally to this work.}
\affiliation{Center for Computational Quantum Physics, Flatiron Institute, New York, NY, 10010, USA}
\author{Vladimir~Calvera} 
\thanks{A.V. and V.C. contributed equally to this work.}
\affiliation{Department of Physics, Stanford University, Stanford, CA 94305, USA}
\affiliation{Kavli Institute for Theoretical Physics, University of California, Santa Barbara, California 93106, USA}
\author{Yubo~Yang}
\affiliation{Center for Computational Quantum Physics, Flatiron Institute, New York, NY, 10010, USA}
\affiliation{Department of Physics and Astronomy, Hofstra University, Hempstead, New York 11549, USA}
\author{Miguel~A.~Morales}
\affiliation{Center for Computational Quantum Physics, Flatiron Institute, New York, NY, 10010, USA}
\author{Steven~A.~Kivelson}
\affiliation{Department of Physics, Stanford University, Stanford, CA 94305, USA}
\author{Ilya~Esterlis}
\affiliation{Department of Physics, University of Wisconsin-Madison, Madison, Wisconsin 53706, USA}
\author{Shiwei Zhang}
\affiliation{Center for Computational Quantum Physics, Flatiron Institute, New York, NY, 10010, USA}

\date{\today}

\begin{abstract}
We report on the properties of the two-dimensional electron gas in a dual-gate geometry, using quantum Monte Carlo methods to obtain aspects of the phase diagram as a function of electron density and gate distance. We identify the critical gate distance below which the Wigner crystal phase disappears. For larger gate distances, the system undergoes a re-entrant transition from crystal to liquid at sufficiently low density. We also present preliminary evidence for a fully polarized ferromagnetic liquid state at low electron density and intermediate gate distances. The quantum Monte Carlo results are compared with simpler approximate methods, which are shown to be semi-quantitatively reliable for determining key features of the phase diagram. These methods are then used to obtain the phase boundary between the  Wigner crystal and liquid in the single-gate geometry. 
\end{abstract}

\maketitle

{\it Introduction.} The two-dimensional electron gas (2DEG) plays a central role in the study of strongly correlated electronic systems, 
largely due to its simplicity and the rich phenomena it exhibits. As a function of increasing density $n$, the 2DEG can be tuned from a solid phase at small $n$ -- the Wigner crystal (WC) -- to a strongly correlated liquid 
and ultimately to a weakly interacting Fermi gas at large $n$. While the detailed nature of the zero-temperature phase diagram as a function of $n$ remains a subject of active investigation \cite{Spivak2004Micro,Jamei2005Coul_Frust,Falakshahi2005_hybrid,*Waintal2006_hybrid,chubukov2017superconductivity,Kim2022Interstitial,Kim2024Self_Dope, smith2024ground,Sung2023,yang2024_dg}, many salient properties of the 2DEG are now quantitatively well understood, thanks primarily to quantum Monte Carlo methods~\cite{Tanatar_GroundStateTwodimensional_1989,Attaccalite20022DEG_QMC,drummond2009phase,smith2024ground}.

In most experimental setups, the 2DEG is created near metallic gate electrodes, separated by a distance $d$ from the electron layer \footnote{For a given experimental system, the value of $d$ may have to be rescaled $d \to \sqrt{\epsilon_\parallel/\epsilon_\perp} \times d$ to properly account for an anisotropic dielectric constant of the 2DEG environment; e.g., as in hBN \cite{Laturia2018}.}. 
The effects of the gate on the properties of the electronic system become significant when $d$ is smaller than the average inter-electron distance $r_s$ (measured in units of the Bohr radius $a_B$), at which point gate screening renders the Coulomb interactions
effectively 
short-ranged. As an additional tuning parameter in the phase diagram of the 2DEG, the gate distance $d$ influences the competition between the electron liquid and WC phases -- causing the WC phase to disappear entirely for sufficiently small $d$ \cite{Spivak2004Micro} -- and may even stabilize new phases not present in the pure Coulomb problem. 

\begin{figure}[b]
\includegraphics[scale=1.0]{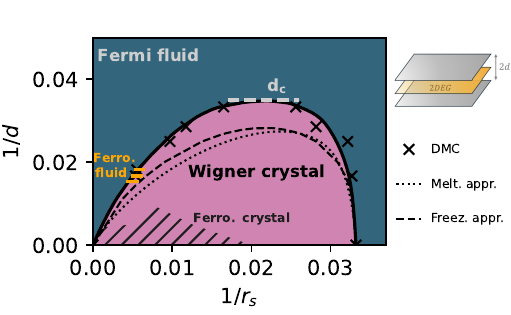}
\caption{The phase diagram of the dual-gate screened two-dimensional electron gas as a function of 
average electron separation $r_s$ and gate distance $d$ (both in units of $a_B$). A schematic of the 
symmetric dual-gate setup is shown in the upper right. Data points of phase boundaries obtained via DMC are depicted as black crosses; the black solid line interpolates the DMC data. The phase boundary between the Wigner crystal and the Fermi fluid obtained via the approximate methods described in the text are shown as a dashed and dotted lines. Hatched regions are
rough estimates of possible regions of stability of ferromagnetic phases. The critical gate distance $d_c \approx 29$ is marked with a dashed grey line. 
Outside of the hatched regions, the default was
paramagnetic for fluid and anti-ferromagnetic for Wigner crystal.
}
\label{fig:fig1}
\end{figure}

In the present paper, we use Diffusion Monte Carlo (DMC) to investigate the effects of symmetric gate electrodes on the properties of the clean 2DEG. We focus on the liquid and WC phases, and map out the liquid-solid phase diagram in the $(r_s, d)$-plane. In particular, we identify the critical gate distance $d_c$ for the existence of a WC phase in the dual-gated geometry. We also provide preliminary evidence for the existence of a narrow region of ferromagnetic liquid at low density and finite $d$. Our primary results are summarized in the phase diagram in Fig.~\ref{fig:fig1}.

We also highlight some approximate methods for determining ground-state energies and the liquid/solid phase boundary that we have found to be in semi-quantitative agreement with the more accurate DMC calculations. 
The reliability of these approximate methods suggests a simple route by which to make estimates for the more structured 2DEGs realized in different materials and devices. 
As an additional example, we analyze the single gate system using these methods.

{\it Screened two-dimensional electron gas.} We consider a 2DEG with interactions screened by two symmetric, parallel metal gates, each
at a distance $d$ from the 2DEG (see Fig.~\ref{fig:fig1}). The Hamiltonian is given by
\begin{align}
\label{eq:H}
H=-\sum \limits_{i} \frac{1}{2} {\bf \nabla}_{i}^2  + \sum \limits_{i<j} V_{\rm sc}(|{\bf r}_i- {\bf r}_j|) + \text{b.g.}.
\end{align}
Labels $i,j$ run over all electrons, $V_{\rm sc}$ is the dual-gate screened potential, and ``b.g." represents the neutralizing background.
Throughout this work we measure lengths in units of the Bohr radius $\aB =  4\pi \epsilon_0 \hbar^2/m e^2$ and energies in Hartree ${\rm Ha} = \hbar^2/m\aB^2$. 

The phase diagram of the dual-gated system is determined by the gate distance $d$ and the Wigner-Seitz radius $r_s = 1/\sqrt{\pi n}$, where $n$  is the electron density. 
On the one hand, the {\it strength} of the interaction is controlled by $r_s$: 
The ratio between (classical) potential and kinetic energy is roughly measured by $r_s$. 
On the other hand, the {\it range} of the interactions is controlled by the gate distance $d$: The metal gates 
screen the interactions which results in strong suppression at distances larger than $d$. The gate-screened interaction is given by
\begin{align}
V_{\rm sc}(|{\bf r}_i-{\bf r}_j|)&=\frac{1}{(2\pi)^2}\int {\rm d} {\bf q}\, {\rm e}^{i{\bf q} \cdot ({\bf r}_i-{\bf r}_j)}v^{}_{\rm sc}({\bf q}), \label{eq:V}\\
v_{\rm sc}({\bf q})&= \frac{e^2}{2 \epsilon_0 } \frac{\tanh (d |{\bf q}|)}{|{\bf q}|},\label{eq:V:dg}
\end{align}
with limiting behavior $V_{\rm sc}(r)\sim{\rm e}^{-\pi r/2d}$ for $r \gg d$ \cite{throckmorton2012fermions}.

\begin{figure}[tb]
\includegraphics[scale=0.93]{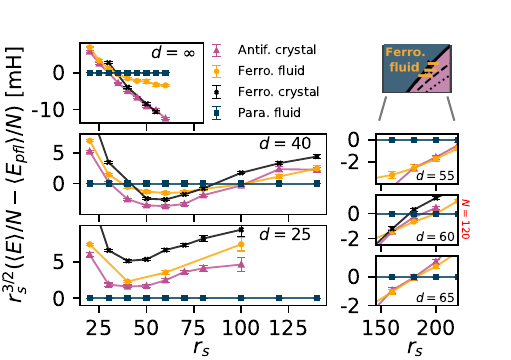}
\caption{
Energies obtained via DMC for selected values of $d$, using $N=56$ electrons. The fluid phases are twist-averaged over $25$ quasi-random twists. On the right hand side, the energetics suggest a possibly stable ferromagnetic fluid. The energies at $d=60$ are computed using $N=120$ in order to further reduce finite-size effects.}
\label{fig:fig2}
\end{figure}

{\it DMC approach.}
DMC constitutes one of the most accurate methods for the study of strongly correlated systems such as the electron gas \cite{ceperley1980ground, foulkes2001quantum, drummond2009phase, Attaccalite20022DEG_QMC, azadi2024quantum}, achieving quantitative agreement with experiments while having no tunable parameters~\cite{Falson_CompetingCorrelatedStates_2022,Sung2023}.
Within DMC, an ensemble of electron configurations is evolved in imaginary time towards the ground state. Fermionic symmetry is ensured by employing the fixed-node (phase) approximation. 
Here, the nodal structure is fixed by a trial wave-function, with parameters optimized by a variational Monte Carlo (VMC) calculation for each system ($r_s$, $d$, number of electrons, and boundary condition).
Here, we utilize trial wave-functions of the Slater-Jastrow and Slater-Jastrow-Backflow type \cite{jastrow1955many, lee1981green, schmidt1981structure, kwon1993effects, kwon1998effects} within the package {\it QMCPACK} \cite{kim2018qmcpack, kent2020qmcpack}, with appropriate modifications. The orbitals in the Slater determinant are evaluated at quasiparticle coordinates if backflow transformations are employed (which we only use for fluid trial wave-functions).
The Jastrow factor and backflow transformation are parametrized using polynomial bsplines \cite{kim2018qmcpack}.

As a result of the fixed-node (phase) approximation, the choice of orbitals reflects the physical phase of the system. Capitalizing on the variational nature of DMC, we study the phase diagram of the screened 2DEG by comparing energies obtained with different physical choices of single-particle orbitals.
Here we consider both fluid and crystalline trial states of varying polarization. 
For the Fermi fluid, we utilize plane wave orbitals $\exp(i {\bf k} \cdot {\bf r})$ to construct paramagnetic and ferromagnetic fluid trial wave-functions. 
For the crystalline states, we use Gaussian orbitals of the form  $\exp(-Cr^2)$, localized on the sites of the triangular WC lattice, with width controlled by the variational parameter $C$. 
We consider 
only  the antiferromagnetic (with alternating lines of spin-up and spin-down electrons) and ferromagnetic crystals, based on results in the unscreened 2DEG \cite{drummond2009phase, smith2024ground}\footnote{We note that more complex anti-ferromagnetic orders have not been ruled out \cite{misguich1998,*misguich1999,Bernu2001}.}.
 All simulations are performed with $N=56$ electrons (and $N=120$ in special cases). Finite-size effects are reduced by performing twist-averaging using $25$ quasi-random twists for all liquid wave-functions \cite{qin2016benchmark}, chosen from the low-discrepancy Halton sequence \cite{halton1960efficiency}.
Many-body finite-size effects due to the interaction remain~\cite{Chiesa_FiniteSizeErrorManyBody_2006,Holzmann_TheoryFiniteSize_2016}, but are expected to be small, especially as the Coulomb potential is screened.  

{\it Extrapolated estimates.} 
 We have also utilized two 
 (computationally inexpensive) approximate approaches to determine the WC-liquid phase boundary in the dual-gated system. We assess the quality of the results by comparing them to the more accurate, but much more computationally intensive, DMC results.

First, the phase transition can be approached from the crystalline side by utilizing the Lindemann melting criterion, which states that melting occurs when the fluctuations of the particles about their equilibrium lattice positions exceed a certain fraction of the lattice spacing. That is, when the parameter $\gamma \equiv \sqrt{\langle{(\delta \rbf_j)^2} / a^2}$ exceeds a critical value $\gamma_c$, where $a$ is the lattice constant. We have computed the fluctuation  $\langle{(\delta \rbf_j)^2}\rangle$ in the harmonic approximation. 
We fix $\gamma_c \approx 0.3$ by matching to the $d=\infty$ transition (the unscreened Coulomb problem) at $r_s \approx 30$.
Details of  the Lindemann-based estimate are presented in  Appendix \ref{app:semiclass}.

Alternatively, we can approach the phase transition from the liquid side and determine the critical values of $r_s$ via the quantum Hansen-Verlet (HV) freezing criterion \cite{babadi2013universal}, which states that the freezing occurs when the maximum of the liquid-state structure factor $S(k)$ attains a critical value $S(k_\text{max})\approx 1.5$. 
To obtain a suitable approximation of the structure factor of the gate-screened 2DEG, we use a variational method \cite{Skinner2010Variational2DEGscreened,babadi2013universal}. We take as a variational wavefunction the ground state of the \textit{unscreened} system with an ``effective" electron charge as a variational parameter, which is then determined by minimizing the energy.
The correlation functions of the unscreened system 
are taken from the parametrization in \cite{GoriGiorgi2004PairDistribution2DEG_MC}.  
The specifics of the HV estimate  are presented in  Appendix \ref{app:var}. 

\begin{figure}[tb]
    \includegraphics[scale=1]{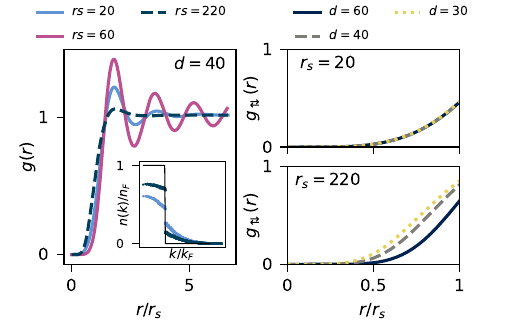}
    \caption{Left: Pair correlation function $g(r)$ at $d=40$, showing Fermi fluids at $r_s=20$ and $r_s=220$, and a Wigner crystal state $r_s=60$. The inset shows the momentum distribution $n(k)$ for the two fluid states. Right: Different short-range behaviors of the spin-resolved correlation function $g_{\uparrow \downarrow}$ in the low-$r_s$ Fermi liquid (upper panel) and the re-entrant Fermi liquid (lower panel), as the gate distance $d$ is varied. } 
    \label{fig:fig3}
\end{figure}

{\it Results (double-gate). }
The phase diagram of the dual-gate screened 2DEG as a function of $r_s$ and $d$, obtained by DMC, is shown in Fig.~\ref{fig:fig1}. The locations of the WC-liquid phase boundaries obtained by the two approximate approaches are also shown for comparison, which are seen to be in qualitative agreement with DMC. We proceed by elaborating upon the DMC simulations and results, followed by a more detailed comparison to the approximate approaches.

We have analyzed the stability of  the crystalline and fluid phases in the regime $20\leq r_s \leq 220$ and $d \geq 20$, chosen such that the screening length and inter-electron distance are of similar 
magnitude. 
The DMC simulations are performed on a finite 
$r_s$-$d$ 
grid. The DMC phase boundary is obtained by comparing the energies of fluid and crystalline states of different spin-polarization and spin-ordering, as shown in Fig.~\ref{fig:fig2} for selected values of $r_s$ and $d$. We have not tested for the possibility of any intermediate partially melted electron liquid crystalline phases \cite{Falakshahi2005_hybrid,*Waintal2006_hybrid,Kim2024Self_Dope}, or partially polarized fluid phases.

As a function of gate distance $d$, three main regimes are observed: (i) In the unscreened limit $d \to \infty$, interactions are long-range and a crystalline phase is stable for all sufficiently low densities (large $r_s$). When the density increases (decreasing $r_s$), the crystalline phase gives way to a (paramagnetic) fluid at about $r_s \approx 30$ \footnote{This value is consistent with the estimated critical $r_s$ in 2DEG using the same method (DMC with SJB trial wave functions)~\cite{drummond2009phase}, as expected. Recent calculations with neural quantum states indicate a  value which is larger~\cite{smith2024ground}, but this is outside the scope of our study.}.
(ii) At finite $d$ with $30 \leq d < \infty $, the region of stability of the Wigner crystal is bounded at {\it both} low and high density. In particular, at sufficiently low density, when the inter-electron distance is larger than the effective interaction range, 
a metallic state is stable.
(iii) For $d<30$, the ground state is a Fermi fluid for all simulated values of $r_s$. We thus identify $d_c \approx 29$ as the critical gate distance for the existence of a WC state; for $d<d_c$ the ground state is liquid for any density. 

\begin{figure}[tb]
    \includegraphics[scale=1]{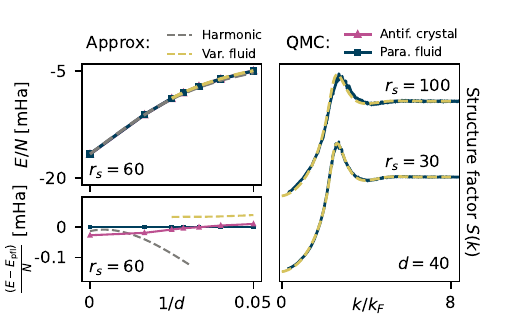}
    \caption{Left: Comparison between approximate and DMC energies. In the upper plot, the absolute energies are compared. Energy differences are resolved in the lower plot, showing the same data but subtracting the DMC energy of the paramagnetic fluid. Right: Structure factor $S(k)$ for the DMC-obtained paramagnetic Fermi fluid and the semiclassical variational fluid, at the two phase boundaries plotted with absolute offset for better visibility.
    }
    \label{fig:fig4}
\end{figure}

The presence of metal gates not only affects the WC-liquid transition, but also influences the spin-ordering and spin-polarization of the system. In the crystal phase, the ferromagnetic state is suppressed for stronger gate screening (smaller $d$) as the interaction becomes more short-ranged. In the liquid phase, our results suggest the possibility of a region of stability for a {\it ferromagnetic fluid} for gate distances $55\lesssim d \lesssim 65$ as presented in Fig.~\ref{fig:fig2}. 
These results are in contrast with the case of a long-range Coulomb interaction, where no region of stability for the ferromagnetic fluid has been observed. 
However, we note that more careful studies are necessary to make a conclusive statement. In particular, the ferromagnetic fluid is in very close competition with the antiferromagnetic WC, with the uncertainty in the energies being of similar 
magnitude 
as the energy difference.

We probe the nature of the ground states further with the pair correlation function $g(r)$, which is shown in Fig.~\ref{fig:fig3} for different values of $r_s$ at a gate distance $d=40$. Long-range oscillations are evident for Wigner crystallization at $r_s=60$, whereas they are damped in the fluid phase at $r_s=20$ and in the re-entrant fluid at $r_s=220$. While the states at $r_s=20$ and $r_s=220$ are both Fermi liquids (as further confirmed by the momentum distribution $n(k)$ in Fig.~\ref{fig:fig3}), subtle differences arise as a consequence of the effective interaction length $\xi\propto d$.
In particular, at low $r_s$ the effective interaction length $\xi\gg r_s$ and thus the shape of the correlation hole is solely determined by the inter-electron distance $r_s$: When plotted at given $r_s$ for different values of gate distance $d$, the short-range behavior of $g_{\uparrow\downarrow}(r)$ does not change -- see
Fig.~\ref{fig:fig3}. On the other hand, 
in the re-entrant phase, where $\xi<r_s$,
the gate distance affects the short-range behavior of $g_{\uparrow\downarrow}(r)$. As depicted in Fig.~\ref{fig:fig3}, the
correlation hole is more pronounced for larger gate distances $d$ (larger $\xi$) if $r_s$ is kept fixed.

The phase boundaries obtained by the Lindemann and HV criteria are in close agreement with each other and in semi-quantitative agreement with DMC (Fig.~\ref{fig:fig1}). In particular, the critical gate distance estimated by the approximate methods is $d_c \approx 35$, exceeding the DMC-obtained value $d_c \approx 29$ by only about 20\%. A direct comparison of ground state energies, shown in Fig.~\ref{fig:fig4}, demonstrates that the qualitative $d$-dependence is accurately captured by both the harmonic approximation utilized for the Lindemann criterion and the variational method used for the HV criterion. However, Fig.~\ref{fig:fig4} also shows that, in contrast to DMC, the accuracy of the approximate methods is not sufficient to resolve the energy differences between different states.  Nevertheless, 
both approximate methods offer sensible routes to estimate observables at negligible computational cost. For example, the structure factor in the fluid phase obtained by the variational method is in good agreement with DMC results (see Fig.~\ref{fig:fig4}). Similarly, the pair correlation function $g(r)$ computed in the crystalline phase via DMC and the harmonic approximation are in reasonable agreement with each other (see App.~\ref{app:gr}).

{\it Single-gate screening.} Another common experimental setup involves a 2DEG near a single metallic gate electrode.
While the dual-gate screened interaction decays exponentially at large distances, the decay is polynomial when there is only a single gate at a distance $d$ from the electron layer, with $V_{\rm sc}(r)\sim d^2/r^3$ for $r \gg d$. Given the reasonable accuracy of the Lindemann melting and HV freezing criteria in the dual-gated case, we have applied them to the single-gate geometry. The results presented in Appendix \ref{app:SingleGate} are qualitatively similar to those in the dual-gate case, with a slight reduction in the critical gate distance, $d_c \approx 27.4$ (Lindemann) and $d_c \approx 25.9$ (HV),  needed to completely suppress the WC. 

\textit{Discussion.} We have analyzed the competition between solid and liquid phases in the phase diagram of the symmetric dual-gated 2DEG using DMC methods.  The WC phase forms a dome in the $(r_s,d)$ plane, consisting of a smaller dome of ferromagnetic crystal within a larger striped antiferromagnetic crystal phase. We have identified the critical gate distance for the existence of the WC as $d_c \approx 29$.  Our results suggest the possibility of a narrow region of ferromagnetic fluid at large $r_s$ and intermediate $d$. The DMC results for the phase boundary were found to be in reasonable agreement with the phase boundaries obtained by the Lindemann melting and HV freezing criteria. These melting and freezing criteria were also used to estimate the solid-liquid phase boundary in the case of a single gate electrode. Our results should be of practical importance for experimental 2DEG systems, which are often realized in gated devices.

Further investigation is required to determine whether additional intermediate phases may emerge or be stabilized by the presence of screening gates. Comparisons of the Lindemann melting and HV freezing criteria with DMC suggest that these approaches should be useful for making estimates in other experimentally realized 2DEG systems, such as those involving more complex gate geometries or dielectric environments.  

~\\

\textit{Acknowledgments.} The Flatiron Institute is a division of the Simons foundation.
This work was supported in part by NSF-BSF award DMR2310312 at Stanford (V.C. and S.A.K.).
This research was supported in part by grant NSF PHY-2309135 to the Kavli Institute for Theoretical Physics (KITP). This research was also supported by the National Science Foundation (NSF) through the University of Wisconsin Materials Research Science and Engineering Center Grant No. DMR-2309000 (I.E.).

\appendix

\def\Nel{N_{\text{el}}}
\def\nel{n_{\text{el}}}
\def\rs{r_{\text{s}}}
\def\rd{r_{\text{d}}}
\def\rbs{\boldsymbol{r}}
\def\rsopt{\rsvar^\star}
\def\Emc{\mathcal{E}}
\def\HH{\mathrm{H}}
\def\FF{\mathrm{F}}
\def\PM{\mathrm{PM}}
\def\FM{\mathrm{FM}}
\def\p{\pi}
\def\vnot{\tilde{v}_0}

\section{Variational calculation using the Coulomb wave-function}
\label{app:var}

This calculation follows Ref.~\cite{Skinner2010Variational2DEGscreened}. Our goal here is to test how good of a variational energy we can get for the liquid states using the variational wave-functions optimized for electrons interacting via pure Coulomb interactions for the problem of electrons interacting via screened Coulomb interactions. 

More concretely, we want to find variational energies for a system with $\nel$ electrons per area interacting via two-body interaction $v(\rbs)$. We use the variational wave function obtained for the same density of electrons interacting via $v_{\text{var}}(\rbs) = \frac{\eevar}{4\pi \epsilon_0 r}$ and optimize over $\eevar$, which is equivalent to optimization over $\rsvar \equiv \frac{m \eevar}{4\pi \epsilon_0\sqrt{\pi \nel}}$.

We used the parametrization of the energy from \cite{Attaccalite20022DEG_QMC} and the parametrization of the PDF from \cite{GoriGiorgi2004PairDistribution2DEG_MC}, which used Monte Carlo simulations for $\Nel =42 (45)$  electrons on a square-shaped box with periodic boundary conditions for the paramagnetic (ferromagnetic) liquid.

\begin{figure}[b]
    \centering
    \includegraphics[width=0.85\linewidth]{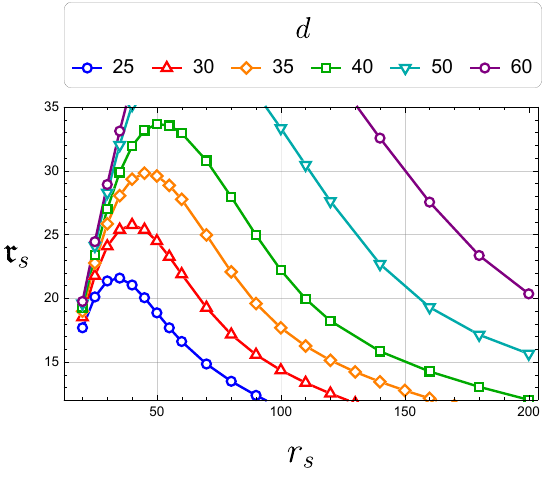}
    \caption{Optimal variational parameter $\mathfrak{r}_s$ as a function of $r_s$ for several values of screening $d$ for the double-gate potential. }
    \label{fig:double_gate_optimalrs}
\end{figure}

We find the optimal value of $\rsvar$ as a function of $\rs = \frac{me^2}{4\pi \epsilon_0\sqrt{\pi \nel} }$ and $ \rd = {\sqrt{\pi \nel d^2}}$. When $\rd$ is large, there is no screening so $\rsvar \approx \rs$. However, when $\rd$ is small the screening is effective so $\rsvar \ll \rs$. In Fig.~\ref{fig:double_gate_optimalrs}, we show a plot of $\rsvar$ versus $\rs$ for various values of $d= {\rs}{\rd}$ and for the double gate cases. 

\section{Semi-classical approximation for Wigner crystal }
\label{app:semiclass}

The semiclassical approximation for the Wigner crystal is not variational. For the Coulomb WC, this approximation works well up to $\rs \approx 30$. For instance, the energy and pair distribution functions match semi-quantitatively with Monte Carlo calculations. 

We modify the classical calculation Ref.~\cite{Bonsall1977_2DWC} for the Coulomb potentials screened by two gates at a distance $d$ from the 2DEG. 

\subsection{Minimization of classical energy}\label{app:ClassicalEnergy}

Deep in the Wigner crystal phase, the electrons minimize the interaction energy by staying as far as possible from each other. We assume that this achieved by localizing the electrons in a Bravais lattice, $\Lambda$. 

The classical energy per electron can be written as 
\begin{equation}
    \frac{E_{\rm Cl}}{\Nel} = \frac{1}{2}\sum_{{\bm R}\in \Lambda: {\bm R}\neq {\bm 0}} V(|{\bm R}|).
\end{equation}
In terms of the basis vectors ${\bm a}_1$ and ${\bm a}_2$, we can write ${\bm R} = n_1 {\bm a}_1 + n_2 {\bm a}_2$ with $n_1,n_2\in \mathbb{Z}$, then the lattice is specified by the inner products $g_{ij}\equiv \langle {\bm a}_i| {\bm a}_j\rangle$. The unit cell size is specified by $\sqrt{|\det(g)|}$ and it is equal to $1/\nel$.

We now restrict to the case of interactions of the form
\begin{equation}\label{eq:V:PosGaussian}
    V(r) = \int_0^{\infty} {\rm e}^{- t r^2}\Phi(t)\dd{t}
\end{equation}
with $\Phi(t) \geq 0$. The potential $V_{\rm{sc}}$ of Eq.~\ref{eq:V:dg} is of this form with 
$\Phi(t) = \frac{1}{\sqrt{\pi t}}\sum_{n\in \mathbb{Z}} (-1)^n {\rm e}^{-4n^2 d^2 t} = \frac{1}{\sqrt{\pi t}}\theta_{4}(e^{-4d^2 t}) =\frac{1}{2d t}\theta_{2}(e^{-\pi^2/(4d^2t)})$. Here $\theta_{2}$ and $\theta_4$ are Jacobi Theta functions.

The classical energy can then be written as 
\begin{equation}
    \frac{E_{\rm Cl}}{\Nel} = \frac{1}{2}\int_0^{\infty} \left[\vartheta_{\Lambda}(t) - 1\right]\Phi(t)\dd{t}
\end{equation}
with 
\begin{equation}
    \begin{split}
        \vartheta_{\Lambda}(t) &= \sum_{{\bm R} \in \Lambda}{\rm e}^{- t |{\bm R}|^2}
    = \sum_{{\bm m} \in \mathbb{Z}^2}{\rm e}^{- t b({\bm m})};\\
    b([m_1,m_2])&= m_1^2 g_{11} + 2g_{12}m_1m_2  + m_2^2 g_{22}.
    \end{split}
\end{equation}

Identifying $t = 2\pi \alpha/\nel$ and $f(m_1,m_2) = (\nel g_{11}) n_1^2 + (\nel g_{22}) n_2^2 + (\nel g_{12}) n_1n_2$ in Theorem 1 in Ref.~\cite{Montgomery1988MinimalTheta}, we deduce that $E_{\rm Cl}$ is minimized when $\Lambda$ is the triangular lattice. 

\subsection{Separation of scales}
For various calculations, it is convenient to split the potential into a long-ranged piece and a short-ranged piece. 

For the potentials of the form in Eq.~\ref{eq:V:PosGaussian}, this can be achieved in terms of a length scale $\sqrt{t_0}$:
\begin{align}
    V_{\rm{LR}}(r;t_0) \equiv \int_{0}^{t_0} {\rm e}^{- t r^2}\Phi(t)\dd{t};\\
    V_{\rm{SR}}(r;t_0) \equiv \int_{t_0}^{\infty} {\rm e}^{- t r^2}\Phi(t)\dd{t}.
\end{align}
LR and SR stand for long-ranged and short-ranged, respectively. One can see that the LR and SR become short-ranged and long-ranged, respectively, in momentum space:
\begin{align}
    \tilde{V}_{\rm{LR}}(q;t_0) = \int_{0}^{t_0} {\rm e}^{- \frac{q^2}{4t}}\frac{\pi\Phi(t)}{t}\dd{t};\\
    \tilde{V}_{\rm{SR}}(q;t_0) = \int_{t_0}^{\infty} {\rm e}^{- \frac{q^2}{4t}}\frac{\pi\Phi(t)}{t}\dd{t}.
\end{align}
where $\tilde{V}_{\rm \beta}(q;t_0) = 2\pi\int_0^{\infty} J_0(qr) V_{\beta}(r;t_0)r\dd{r}$ with $\beta={\rm SR, LR}$ and $J_0$ is the zeroth-order Bessel function. 

For the Coulomb potential, it is convenient to choose $t_0 = \pi\nel $. In this case
\begin{align}
V_{\rm SR}(r; \pi \nel) &= 
\frac{\erfc(\sqrt{\pi \nel r^2})}{r}\\
V_{\rm LR}(r; \pi \nel) &= 
\frac{\erf(\sqrt{\pi \nel r^2})}{r}\\
\tilde{V}_{\rm SR}(q; \pi \nel) &= 
\frac{2\pi \erf(\sqrt{ q^2/(4\pi \nel )})}{q}\\
\tilde{V}_{\rm LR}(q; \pi \nel) &= 
\frac{2\pi \erfc(\sqrt{ q^2/(4\pi \nel )})}{q}
\end{align}

For the double gated potential and poor screening ($\pi \nel d^2\gtrsim 1$), it is convenient to split
\begin{equation}\label{eq:Scales:large_rd}
    \begin{split}
    {V}_{\rm SR}(r) &= \frac{1}{r} \erfc(\sqrt{ \pi \nel r^2 }),\\
        \tilde{V}_{\rm LR}(q) &= \frac{2\pi}{q}\left( \erfc(\sqrt{ q^2/(4\pi\nel) })- \frac{2}{e^{2 qd }+1} \right).
    \end{split}
\end{equation}
In other words, we use separation of scales for the direct piece of the interactions. The sums over the interaction with the image charges is done in momentum space because $2/(e^{2qd}+1)$ decays exponentially. 

When screening is good ($\pi \nel d^2\lesssim 1$), we evaluated the sums directly in real space because the potential decays exponentially as $e^{- \pi r/2d}$. To be more precise, we use the expression $V_{\rm{sc}}(r) = \frac{e^2}{4\pi \epsilon_0 r} \upsilon(\tfrac{\pi r}{2d})$, where
\begin{equation}
    \upsilon(x) = \int_0^{\infty} \frac{2}{\pi}\frac{x}{\sinh(\cosh(s) x)}\dd{s}.
\end{equation}
The latter expression for $V_{\rm{sc}}$ is obtained by combining Eq.~{B7} of Ref.~\cite{throckmorton2012fermions} with the integral representation for the modified Bessel function $K_0(x) = \int_0^{\infty}e^{- x \cosh(s)}ds$.

\subsection{Evaluation of the classical energy}
\def\Rbs{{\bm R}}
\def\Gbs{{\bm G}}
\def\zero{{\bm 0}}

\begin{figure}[h!]
    \centering
    \includegraphics[width=0.95\linewidth]{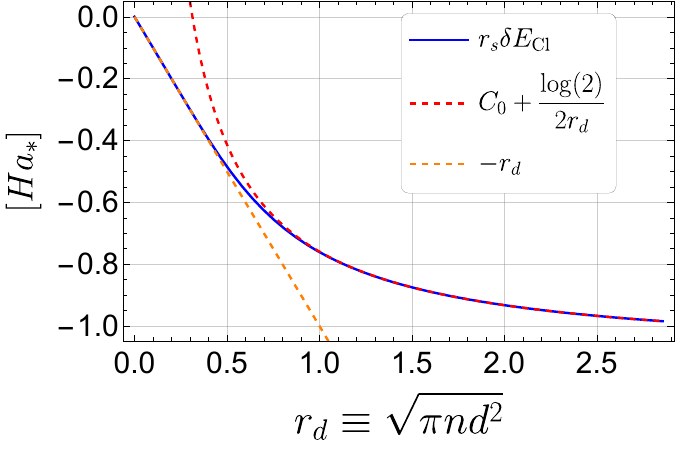}
    \caption{Classical energy times $\rs$ versus screening parameter $r_d = \sqrt{\pi \nel d^2}$. $C_0 \approx -1.1061$ is the value for pure Coulomb interactions. The correction to the dashed lines are exponentially small in $\rd$ or $1/\rd$. }
    \label{fig:EnergyWC:Classical}
\end{figure}

To compare the energy of the crystal with the energy of the fluid, it is convenient to split the uniform component of the energy. 
\begin{align}
    \frac{2E_{\rm Cl}}{\Nel} &= \nel \tilde{V}(0) + U_{\rm SR}+ U_{\rm LR} ; \label{eq:ECl:WC}\\
    U_{\rm SR} &=-V_{\rm LR}(0)+ \sum_{\Rbs \in \Lambda}'V_{\rm SR}(|\Rbs|);
    \\
    \frac{U_{\rm LR}}{\nel} &=-\tilde{V}_{\rm SR}(0)+ \sum_{\Gbs \in \Lambda^{\vee}}'\tilde{V}_{\rm LR}(|\Gbs|),
\end{align}
where $\Lambda^\vee$ is the dual lattice and the primed sums mean to omit the zero vector. The first term 
in Eq.~\ref{eq:ECl:WC} is canceled by the neutralizing background.
We thus identify the classical energy 
per electron for the crystal phase in the presence of a neutralizing background as 
\begin{equation}
    \delta E_{\rm Cl}\equiv \frac{U_{\rm SR} + U_{\rm LR}}{2}. 
\end{equation}
In Fig.~\ref{fig:EnergyWC:Classical}, we show $\delta E_{\rm Cl}$ for the double-gate potential as a function of $\rd\equiv \sqrt{\pi n d^2} = d/\rs$.
The answer for the Coulomb case is $C_0^*/a_0$ with $C_0^* \approx -2.10671$ and $a_0$ the WC lattice constant, or equivalently the Madelung constant $C_0 \approx -1.1061$. We obtained the asymptotic behavior as follows. For the large $\rd$, we use the separation of scales in Eq.~\ref{eq:Scales:large_rd}. Then the difference of $\delta E_{\rm Cl}$ with respect to the pure Coulomb case, comes from the terms involving $V_{\rm LR}$. The change from $-V_{\rm LR}(0)$ is 
\begin{equation}
    \int_{0}^{\infty}\, \frac{2\pi}{q} \frac{2}{e^{2qd}+1}  \frac{q\dd{q}}{2\pi} = \log(2)/d
\end{equation}
The other corrections come from the sum at finite momenta in $U_{\rm LR}$. All these contributions are suppressed by $\exp(- 2|{\Gbs}|d)$:
\begin{equation}
    \rs *\delta E_{\rm Cl} = C_0 + \frac{\log(2)}{2\rd} - \frac{1}{\pi} \sum_{\Gbs\in \Lambda^\vee_{\times}} \frac{2\pi}{\rs G}\frac{1}{1+e^{2\rd (\rs G)}}
\end{equation}

For small $\rd$ the interactions decay exponentially, so that to exponential order $\delta E_{\rm Cl}$ is $-\nel \tilde{v}(\zero)/2 = - \pi\nel d = - \rd/\rs$. We get
\begin{equation}
    \rs *\delta E_{\rm Cl} = 
    -\rd + \sum_{\Rbs \in \Lambda_\times} \frac{\upsilon( \tfrac{\pi}{2\rd} \tfrac{R}{\rs})}{2R/\rs}
\end{equation}

\subsection{Phonon spectrum}\label{app:PhononSpectrum}
\def\pbs{{\bm p}}
\def\qbs{{\bm q}}
\def\ee{{\rm e}}
\def\ii{{\rm i}}
\def\Lind{\gamma}
\def\aWC{a}
Now that we have minimized the potential energy, we need to minimize the kinetic energy. To do so, we let the electrons deviate from their equilibrium positions. These oscillations impose a restoring force on the electrons that is of the order $\partial_{R_a} \partial_{R_b} V(R)$. The balance between the restoring force and the kinetic energy gives rise to the collective phonon modes of the Wigner crystal. 

We assume that electrons are well-localized near the sites of the triangular lattice so that we can treat them as distinguishable particles. 
In the harmonic approximation, we obtain 
\begin{equation}
    H_{\rm Har} = \sum_{j} \frac{\pbs^2_j}{2m} + \sum_{i,j} \frac{\delta\rbs_i K_{ij} \delta\rbs_j}{2}
\end{equation}
where $\delta\rbs_j = \rbs_j -\Rbs_j$ is the position of the $j$-th electron relative to its equilibrium position, $\Rbs_j$. The elastic matrix is
\begin{equation}
    \begin{split}
        [K_{ii}]_{ab} &= \sum_{\Rbs}{'} \frac{\partial^2V(R)}{\partial R_a \partial  R_b};\\
        [K_{ij}]_{ab} &=-\frac{\partial^2V(R=|\Rbs_{ij}|)}{\partial R_a \partial  R_b}\quad (j\neq i).
    \end{split}
\end{equation}

The Hamiltonian $H_{\rm Har}$ is diagonalized through a Fourier transform: $\tilde{\pbs}_{\qbs} = \frac{1}{\Nel}\sum_{j} \ee^{-\ii \qbs \Rbs_j} \pbs_j$, 
$\delta\tilde{\rbs}_{\qbs} = \frac{1}{\Nel}\sum_{j} \ee^{\ii \qbs \Rbs_j} \delta\rbs_j$ and
\begin{equation}
    \tilde{K}(\qbs) = \sum_{j\neq 0}\ee^{\ii\qbs\Rbs_j } K_{0j} = \sum_{\Rbs}{'}(1-\cos(\qbs\Rbs)) \kappa(\Rbs),
\end{equation} 
with 
\begin{equation}
    \kappa(\Rbs)_{ab}  = \frac{\mathcal{D}_{\perp}(\delta_{ab} - \hat{R}_a\hat{R}_b) + \mathcal{D}_{\parallel} \hat{R}_a \hat{R}_b}{R^3},
\end{equation}
where $\mathcal{D}_{\perp}$ and $\mathcal{D}_{\perp}$ are functions of the ratio $|\Rbs|/d$.

The dynamical matrix is then $D(\qbs) = \tilde{K}(\qbs)/m$ so that the phonon frequencies $\omega_{k}(\qbs)$ are the (non-negative) square roots of the eigenvalues of $D(\qbs)$, and $k=1,2$ label the two phonon branches.  Recall that $D(\qbs) = (1/\rs^3)\bar{K}(\qbs) $.

The semi-classical energy is then 
\begin{equation}
    \frac{E_{\rm SemiCl}}{\Nel} = 
    \int_{\qbs}
    \frac{\omega_1(\qbs)+\omega_2(\qbs)}{2}.
\end{equation}
where the integral $\int_{\qbs}$ is over the Brillouin zone of the WC and normalized to $\int_{\qbs}1 = 1$ \footnote{In practice, we approximate $\int_{\qbs}$ by a finite sum over a uniform grid of the Wigner crystal' Brillouin zone: $\int_{\qbs} = \frac{1}{4N_1 N_2} \sum_{\qbs}$, with $\qbs = \frac{j_1}{N_1}\Gbs_1+\frac{j_2}{N_2}\Gbs_2$ with $-N_a<j_a\leq N_a \, (a=1,2)$. When calculating the Lindemann parameter we simply omit the $\qbs = \zero$.}.



\subsection{Melting of Wigner crystal}\label{app:MeltingDetails}

We estimate the location of the melting instability through the Lindemann criterion. First, we calculate the Lindemann parameter $\Lind \equiv \sqrt{{\langle \delta\rbs^2_0\rangle}/{\aWC^2}}$, where $\aWC$ is the Wigner crystal lattice constant which is given by $\aWC = \left(\frac{4}{3}\right)^{1/4}/\sqrt{\nel}$ for the triangular lattice. The expectation value should be calculated with the ground state of the (full) phonon Hamiltonian. However, we will use the ground state of the harmonic Hamiltonian, $H_{\rm Har}$. 

We can evaluate $\Lind$ in terms of the phonon spectrum:
\begin{equation}
    \Lind^2 = \frac{1}{\aWC^2}
    \int_{\qbs} \sum_{k} \frac{1}{2m\omega_k(\qbs)}\,.
\end{equation}

For potentials of the form $v(r) = \frac{e^2}{4\pi \epsilon r} u(r/d)$ for some function $u$, we can write $\omega_k(\qbs) = \frac{1}{m \sqrt{\aWC^3\aB}} \bar{\omega}_{k}(\qbs \aWC; d/\aWC)$, with $\bar{\omega}$ dimensionless frequencies that only depend on the shown variables. From this follows that 
\begin{equation}\label{eq:Lind:Scaling}
    \Lind^2 =\frac{\Upsilon(d/\aWC)}{\sqrt{\rs}},\quad \Upsilon = 
    \mathcal{N}\int_{\qbs}\sum_{k} \frac{1}{\bar{\omega}_k(\qbs)},
\end{equation}
with $\mathcal{N} = \sqrt[4]{\frac{\sqrt{3}}{32\pi}} \approx 0.362297$ and $k=1,2$ refers to the two phonon bands. 

Deep in the WC phase, $\Lind\ll 1$. The transition occurs at some critical value $\Lind_c$. This value seems to be `universal' \cite{babadi2013universal}. In practice, we determine $\Lind_{c}$ by calculating $\Lind$ at a point in parameter space we believe the melting transition occurs. From the scaling of the $\Lind$ in Eq.~\ref{eq:Lind:Scaling}, we find the following scaling for the critical $\rs$ at gate distance $d$, $\rs^{\star}(d)$:
\begin{equation}\label{eq:rs:LambdaScaling}
    \frac{\rs^{\star}(d_1)}{\rs^{\star}(d_2)} = \left(\frac{\Upsilon(d_1/\aWC)}{\Upsilon(d_2/\aWC)}\right)^2.
\end{equation}
In the main text, we used that $\rs^\star(\infty) = 30.1$, which is extracted from the transition between the WC and fluid phases. 





\def\kbs{{\bf{k}}}
\def\ubs{{\bf{u}}}

\subsection{Pair distribution function}
\label{app:gr}

We calculate the pair distribution function (PDF) from the phonon spectrum to compare with the Monte Carlo results. 

The PDF is defined to be 
\begin{equation}
    \nel^2 g(\rbs) :=  \frac{\nel}{\Nel}\sum_{i\neq j} \langle{ \delta(\rbs- \rbs_i + \rbs_j)}\rangle,
\end{equation}
where the sum is over electrons, $\nel$ is the electron density, $\Nel$ is the number of electrons and $\rbs_i$ is the position of the electron localized near $\Rbs_i$. The expectation value is taken with respect to the ground state of the harmonic Hamiltonian. 

Using the integral representation of the two-dimensional delta function, we obtain 
\begin{equation}
     \begin{split}
         g(\rbs) 
         &=  \frac{1}{\Nel\nel}\sum_{i\neq j}\int\!\frac{\dd^2{\qbs}}{(2\pi)^2} \langle{ \exp(\ii\qbs\cdot(\rbs- \rbs_{ij}))}\rangle ; \\
         &=  \frac{1}{\Nel\nel}\sum_{i\neq j}\int\!\frac{\dd^2{\qbs}}{(2\pi)^2} \ee^{\ii\qbs\cdot(\rbs-\Rbs_{ij})}
         \ee^{-\frac{\langle{(\qbs\cdot(\rbs_{ij}-\Rbs_{ij}))^2}\rangle}{2}}
         ; 
     \end{split}
\end{equation}
here $\rbs_{ij}:=\rbs_{i}-\rbs_{j}$ and $\Rbs_{ij}:=\Rbs_{i}-\Rbs_{j}$. 

\begin{figure}[t]
    \centering
    \includegraphics[scale=1]{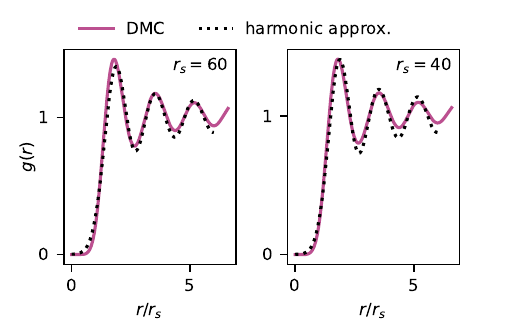}
    \caption{The pair correlation function $g(r)$ computed with DMC and computed from the phonon spectrum (harmonic approximation) at gate distance $d=40$.}
    \label{fig:g_r_dmc_harmonic}
\end{figure}

The expectation value in the exponent can be expressed in terms of the harmonic spectrum as 
\begin{equation}
    \begin{split}
        \left[\mathcal{G}_{ij}\right]_{ab}&:=\langle [\rbs_{ij} - \Rbs_{ij}]_a [\rbs_{ij} - \Rbs_{ij}]_b \rangle \\
        &= \frac{1}{\Nel}\sum_{\kbs,\lambda} \frac{1-\cos(\kbs\cdot\Rbs_{ij})}{m \omega_{\lambda}(\kbs)} [e_{\kbs,\lambda}]_a[e_{\kbs,\lambda}]_b,
    \end{split}
\end{equation}
where $e_{\kbs,\lambda}$ is a real vector satisfying $\tilde{K}(\kbs) e_{\kbs,\lambda} = m [\omega_{\lambda}(\kbs)]^2 e_{\kbs,\lambda}$. When 
$|\Rbs_{ij}|\to\infty$, $\mathcal{G}_{ij}  \sim \delta_{ab}\langle{(\rbs_{0}-\Rbs_{0})2}\rangle$.

From lattice translation invariance, we find
\begin{equation}\label{eq:PDF}
    \begin{split}
        g(\rbs) 
        &= \frac{1}{\nel}\sum_{j\neq 0}\int \frac{\dd^2{\qbs}}{(2\pi)^2} \ee^{\ii\qbs(\rbs-\Rbs_{j0})}\exp(-\qbs\cdot\mathcal{G}_{j0}\cdot\qbs/2) ; \\
        &= \frac{1}{\nel}\sum_{j\neq 0}\frac{\exp( - (\rbs-\Rbs_{j0})\cdot \mathcal{G}_{j0}^{-1}\cdot(\rbs-\Rbs_{j0})/2)}{2\pi\sqrt{\det(\mathcal{G}_{j0})}} .
    \end{split}
\end{equation}

The radial pair distribution function (rPDF) is defined as
\begin{equation}
    \begin{split}
        \bar{g}(r) 
        &:= \int \frac{\dd{\theta}}{2\pi} g(\rbs)
        =\frac{1}{2\pi r} \frac{1}{\nel \Nel}\sum_{i\neq j } \langle \delta( r - r_{ij}) \rangle,
    \end{split}
\end{equation}
where $r_{ij} := |\rbs_{i}-{\rbs}_j|$. We find $\bar{g}(r)$ by numerically integrating  $g(\rbs)$ in Eq.~\ref{eq:PDF}.  

Figure~\ref{fig:g_r_dmc_harmonic} compares the pair correlation function obtained via DMC with the harmonic approximation for the gate distance $d=40$.

\section{Single gate potential}
\label{app:SingleGate}

\begin{figure}[h]
\includegraphics[scale=0.8]{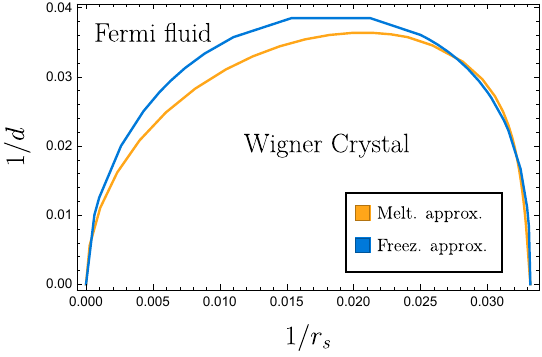}
\caption{The phase diagram of the single-gate screened two-dimensional electron gas as a function of 
average electron separation $r_s$ and gate distance $d$ (both in units of $a_B$). 
}
\label{fig:figSingleGate}
\end{figure}

In the presence of a single gate at a distance $d$, $V_{\rm{sc}}(r) = \frac{e^2}{4\pi\epsilon}\left(\frac{1}{r} - \frac{1}{\sqrt{r^2+4d^2}}\right)$. We show the phase diagram in Fig.~\ref{fig:figSingleGate}
using the freezing approximation (V.H. criteria) and the melting approximation (Lindemann criteria). The approximations do not explicitly account for the magnetic order. However, we anticipate that the Wigner crystal is a striped anti-ferromagnet, while the Fermi fluid is likely unpolarized, as in the double-gated two-dimensional electron gas.

The evaluation of the VH criteria is straightforward. To calculate the Lindemann parameter, we followed the same steps as in Appendices~\ref{app:PhononSpectrum} and ~\ref{app:MeltingDetails}. The main difference is that in the good screening regime, the evaluation in real space is somewhat slow. We instead used the separation of scales (Ewald method) as we did for the pure Coulomb interaction with $\Phi(t)= \frac{1- \exp(-4d^2 t)}{\sqrt{\pi t}}$.

\bibliographystyle{apsrev4-2}
\bibliography{References}

\end{document}